\documentclass{article}

\PassOptionsToPackage{numbers}{natbib}
\usepackage[preprint]{neurips_2025}

\usepackage{amsmath}
\usepackage[utf8]{inputenc} 
\usepackage[T1]{fontenc}    
\usepackage{hyperref}       
\usepackage{url}            
\usepackage{booktabs}       
\usepackage{amsfonts}       
\usepackage{nicefrac}       
\usepackage{microtype}      
\usepackage{xcolor}         
\usepackage{graphicx}

\title{IMPACT: Industrial Machine Perception via Acoustic Cognitive Transformer}

\author{
  Changheon Han\textsuperscript{1} \quad
  Yuseop Sim\textsuperscript{1} \quad
  Hoin Jung\textsuperscript{2} \quad
  Jiho Lee\textsuperscript{1} \quad
  Hojun Lee\textsuperscript{1} \quad 
  Yunseok Kang\textsuperscript{3} \\
  Sucheol Woo\textsuperscript{4} \quad
  Garam Kim\textsuperscript{5} \quad
  Hyung Wook Park\textsuperscript{3} \quad
  Martin Byung-Guk Jun\textsuperscript{1}\textsuperscript{*} \\
  \\
  \textsuperscript{1}School of Mechanical Engineering, Purdue University \\
  \textsuperscript{2}Elmore Family School of Electrical and Computer Engineering, Purdue University \\
  \textsuperscript{3}Department of Mechanical Engineering, UNIST \\
  \textsuperscript{4}Polytechnic Institute, Purdue University \\
  \textsuperscript{5}The School of Aviation and Transportation Technology, Purdue University \\
  \texttt{\{han711, sim46, jung414, lee4503, lee1764\}@purdue.edu}, \\
  \texttt{yskang@unist.ac.kr}, \texttt{\{woo77, kim1652\}@purdue.edu}, \\
  \texttt{hwpark@unist.ac.kr}, \texttt{mbgjun@purdue.edu}
}

\begin{document}

\maketitle



\begin{abstract} 
    Acoustic signals from industrial machines offer valuable insights for anomaly detection, predictive maintenance, and operational efficiency enhancement. However, existing task-specific, supervised learning methods often scale poorly and fail to generalize across diverse industrial scenarios, whose acoustic characteristics are distinct from general audio. Furthermore, the scarcity of accessible, large-scale datasets and pretrained models tailored for industrial audio impedes community-driven research and benchmarking. To address these challenges, we introduce \textbf{DINOS} (\textbf{D}iverse \textbf{IN}dustrial \textbf{O}peration \textbf{S}ounds), a large-scale open-access dataset. DINOS comprises over 74,149 audio samples (exceeding 1,093 hours) collected from various industrial acoustic scenarios. We also present \textbf{IMPACT} (\textbf{I}ndustrial \textbf{M}achine \textbf{P}erception via \textbf{A}coustic \textbf{C}ognitive \textbf{T}ransformer), a novel foundation model for industrial machine sound analysis. IMPACT is pretrained on DINOS in a self-supervised manner. By jointly optimizing utterance and frame-level losses, it captures both global semantics and fine-grained temporal structures. This makes its representations suitable for efficient fine-tuning on various industrial downstream tasks with minimal labeled data. Comprehensive benchmarking across 30 distinct downstream tasks (spanning four machine types) demonstrates that IMPACT outperforms existing models on 24 tasks, establishing its superior effectiveness and robustness, while providing a new performance benchmark for future research. The code and datasets are publicly available at \textcolor{cyan}{\url{https://github.com/hanprd/IMPACT}} and \textcolor{cyan}{\url{https://doi.org/10.7910/DVN/FWAZBQ}}, respectively.
\end{abstract}

\section{Introduction}

\begin{figure}[t]
            \centering
            \includegraphics[width=1.0\linewidth]{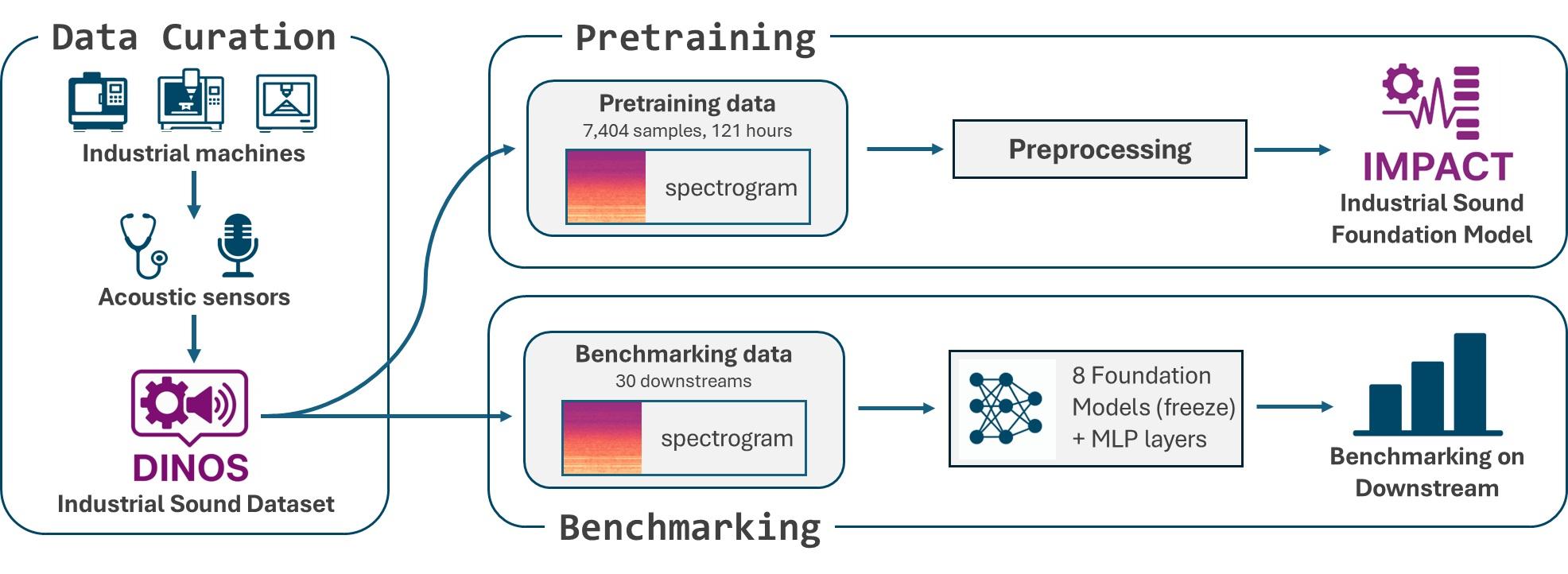}
            \caption{\textbf{System Overview.} After curating data using stethoscope and microphone sensors on industrial machines, we pretrain IMPACT, a transformer-based industrial sound foundation model. Benchmarking evaluates eight sound foundation models on 30 industrial downstream tasks.}
            \label{fig:overview}
            \end{figure}    

Acoustic signals generated by industrial equipment carry critical operational insights for anomaly detection, predictive maintenance, and process optimization, which are essential components for improving system reliability and operational efficiency in manufacturing environments \cite{lee2024stethoscope}. Despite their relevance, most current approaches in industrial settings are task-specific and supervised models, requiring domain-specific labeling efforts, which limits their scalability. Furthermore, publicly available datasets and pretrained models targeting industrial sound are insufficient, limiting reproducibility and slowing progress. Foundation models, which have recently achieved notable success in domains such as natural language processing \cite{devlin2019bert,brown2020language} and computer vision \cite{dosovitskiy2020image,radford2021learning,kirillov2023segment}, provide a potential avenue to overcome these challenges. Leveraging large-scale self-supervised learning, foundation models can learn versatile representations that transfer effectively across multiple downstream tasks with minimal labeled data requirements \cite{zhang2024towards}. 

In the audio domain, similar approaches have emerged \cite{hershey2017cnn,huang2022masked,elizalde2023clap}, yet they primarily focus on general audio like music, YouTube videos, and conversation speech. However, industrial sounds differ from general audio domains like speech or music. These often include stable tonal harmonics tied to machine kinematics (e.g., rotational speeds), specific broadband noise profiles generated by physical processes (e.g., flow, friction), and unique temporal structures such as operational periodicity or diagnostically significant transients arising from faults \cite{randall2021vibration,bies2023engineering,antoni2007cyclic}. As a result, models trained on general-purpose audio fail to capture the acoustic features of industrial environments. To address these challenges, this paper presents two resources intended to shift this landscape (Figure \ref{fig:overview}). 

First, we release \textbf{DINOS} (\textbf{D}iverse \textbf{IN}dustrial \textbf{O}peration \textbf{S}ounds)), which is a large-scale open-access dataset covering a wide range of acoustic scenarios typical in manufacturing contexts. It consists of over 1,093 hours and 74,149 clips across diverse manufacturing processes (e.g., Computer Numerical Control (CNC), Directed Energy Deposition (DED), Laser Powder Bed Fusion (LPBF)). In contrast to earlier datasets (e.g., MIMII \cite{purohit2019mimii}, ToyADMOS \cite{koizumi2019toyadmos}), which focus on limited machine types or controlled faults, DINOS reflects operational diversity and real-world conditions. Employing two types of sensors—microphone and stethoscope sensors, we capture both distinctive and blended acoustic characteristics of machines.

Second, we introduce \textbf{IMPACT} (\textbf{I}ndustrial \textbf{M}achine \textbf{P}erception via \textbf{A}coustic \textbf{C}ognitive \textbf{T}ransformer), the first foundation model specifically designed for industrial machine sound. Pretraining on DINOS enables it to learn industrial acoustic representations beyond the reach of models trained on speech or general environmental sounds. Additionally, we present benchmarking results of multiple pretrained sound foundation models and a fine-tuned model on DINOS to support reproducible research.

Our contributions are summarized as follows:
\begin{itemize}
    \item We release DINOS, a large-scale dataset encompassing CNC, DED, and metal additive manufacturing sounds, explicitly curated for industrial acoustic analysis.
\end{itemize}
\begin{itemize}
    \item We introduce IMPACT, a scalable acoustic foundation model trained via self-supervised methods on industrial machine sound.
\end{itemize}
\begin{itemize}
    \item We empirically validate IMPACT’s superior performance across a range of downstream tasks compared to various sound foundation models, establishing a new benchmark for future research.
\end{itemize}

The remainder of the paper is structured as follows: Section 2 reviews related work, Section 3 details our dataset collection, Section 4 model training methodology, Section 5 presents benchmarking results, and Section 6 concludes with discussions on implications and future directions.

\section{Related Work}
\subsection{Public Audio Datasets}
\paragraph{General-Purpose Audio Corpora.} Research on general acoustic analysis has been supported by a series of publicly available datasets. AudioSet \cite{gemmeke2017audio} remains the most influential, containing over two million human-labeled 10-second sound clips across 527 sound event categories sourced from YouTube. ESC-50 and UrbanSound8K \cite{piczak2015environmental,salamon2014dataset} serve as curated benchmarks for environmental sound classification, with labeled clips drawn from various common audio scenes. For instance, UrbanSound8K comprises 8,732 excerpted urban sounds (<=4s) with 10 classes. These datasets have catalyzed progress in general-purpose acoustic event detection and classification. However, their data reflects common everyday domains such as speech, music, and ambient city noise, and lacks the distinctive acoustic signatures found in industrial settings.

\paragraph{Industrial Machine Sound.} The industrial sound domain has suffered from a lack of open datasets. This was first addressed in 2019 with the release of MIMII \cite{purohit2019mimii} and ToyADMOS \cite{koizumi2019toyadmos}. MIMII provides recordings via a circular microphone array for four machine types (valves, pumps, fans, and slide rails) under both normal and anomalous conditions. ToyADMOS complements this by simulating mechanical faults in miniature machines to generate a large volume of labeled abnormal sounds. These datasets enabled the emergence of benchmark-driven research in Anomalous Sound Detection (ASD). Extensions such as MIMII-DG \cite{dohi2022mimii} introduced domain shift scenarios—e.g., changes in operating conditions, mixing background noise—to study model robustness under varying conditions. While impactful, these datasets still cover a narrow subset of machine types, and their anomalies are often artificially induced or limited in diversity. No prior dataset has captured high-fidelity, real-world machine sound across a diverse set of manufacturing processes with sufficient scale for pretraining.

\subsection{Audio Foundation Models}
\paragraph{General Models.} The transition from hand-crafted features to deep  learning in sound followed patterns seen in vision and language. Early deep models, such as CNN-based architectures like VGG \cite{hershey2017cnn} and PANNs \cite{kong2020panns}, were pretrained on large audio datasets and applied to downstream classification tasks. With the rise of transformer architectures, attention-based models like Audio Spectrogram Transformer (AST) \cite{gong2021ast} emerged. AST demonstrated that patch-based spectrogram tokenization and attention mechanisms could outperform CNNs on ESC-50 and AudioSet, achieving 95.6\% accuracy and 0.485 mAP, respectively. AudioMAE \cite{huang2022masked} further improved general audio representation learning by introducing a masked spectrogram autoencoding objective, showing strong transfer across tasks. In speech processing, self-supervised methods like wav2vec 2.0 \cite{baevski2020wav2vec} and HuBERT \cite{hsu2021hubert} pretrained on thousands of hours unlabeled speech and achieved state-of-the-art performance using minimal supervision. These frameworks demonstrated that context prediction, latent target regression, and other pretext tasks can yield robust audio embeddings. Despite this progress, most models are trained on web-sourced or human-centric audio, which differs from industrial sounds.
\paragraph{Domain-Specific Models.} Motivated by the limitations of general models, several works have developed foundation models for specialized audio domains. For instance, OPERA \cite{zhang2024towards} introduced a model pretrained on 400 hours of respiratory audio from coughs and breathing events. OPERA outperformed general-purpose models on 16 out of 19 medical acoustic tasks, demonstrating that tailoring model training to a specific audio domain yields clear benefits. Other efforts have appeared in areas such as ecology and environment \cite{chasmai2024inaturalist,piczak2015esc}. However, no prior foundation model has been built or trained for the industrial acoustic tasks. Current machine listening models have either relied on features extracted from general audio encoders or been trained from scratch for specific tasks under supervision. Our work fills this gap by proposing DINOS and IMPACT.

\subsection{AI for Industrial Sound Analysis}
\paragraph{Unsupervised Methods.} Anomalous Sound Detection (ASD) has been the primary topic in industrial machine listening. Early methods used hand-crafted spectral features such as Mel-Frequency Cepstral Coefficients (MFCCs) combined with Gaussian Mixture Models (GMMs) or one-class Support Vector Machines (SVMs) \cite{chu2009environmental,sivasankaran2013robust,heittola2013context}. With the advent of deep learning, autoencoders became a dominant framework. \citet{marchi2017deep} showed that deep recurrent autoencoders could capture normal machine operation patterns and represent deviations via reconstruction error. This paradigm was widely adopted due to its independence from labeled anomaly data, which is rare in practice. Most models are trained per machine, assuming that normal sound patterns are device-specific. Consequently, they often fail to generalize across machines or to new operating conditions.
\paragraph{Supervised and Semi-Supervised Approaches.} Where labeled fault types are available, supervised learning has been applied to tasks such as tool wear detection \cite{yun2023autoencoder} and additive manufacturing diagnostics \cite{lee2024stethoscope}. These models often show high accuracy within the training domain but require significant annotation effort and struggle to scale. Semi-supervised approaches attempt to mitigate label scarcity by incorporating limited labeled data into primarily unsupervised training pipelines \cite{han2024visual}. Nevertheless, both supervised and Semi-supervised methods remain impractical in most manufacturing scenarios due to the difficulty of capturing and labeling rare fault events.
\paragraph{Domain Generalization.} A major obstacle to deploying machine listening systems is the issue of domain shift. Models trained on one set of conditions often perform poorly when exposed to unseen environments. The MIMII-DG dataset \cite{dohi2022mimii} was designed to test this explicitly, revealing that even small shifts in machine load or background acoustics can degrade performance significantly. This has spurred interest in domain adaptation techniques, including adversarial feature alignment, feature normalization, and test-time adaptation. Yet, robust generalization remains an open challenge, as most systems are optimized for controlled conditions.

\subsection{Limitations and Contributions}
In summary, the field faces three pressing limitations. First, existing industrial sound datasets are narrow in scope, limited in scale, and insufficient for training foundation models. Second, no publicly available foundation model exists for industrial machine sound, leaving researchers to rely on task-specific solutions or general audio embeddings ill-suited to the domain. Third, current methods often fail to generalize across machines or environments due to strong domain dependence. In response to challenges, we present two contributions: (1) DINOS, a large-scale benchmark dataset of 74,149 sound clips and over 1,093 hours from diverse machines, and (2) IMPACT pretrained via self-supervised learning on DINOS. Our experiments show that IMPACT yields robust, transferable representations that improve performance across various downstream tasks. Together, these contributions aim to establish a scalable foundation for industrial machine listening, enabling broader adoption and reproducible research.

\section{DINOS - Dataset Construction}
\subsection{Data Acquisition}
We collect industrial machine sound data using two sensor types: a USB microphone (K053, Fifine) and a customized stethoscope sensor (K053, Fifine + Dual Head, MDF Instruments). The microphone, directly attached to machines, captures both machine sounds and surrounding noise, which in confined industrial spaces often causes reverberation and crosstalk. In contrast, the stethoscope sensor, with its bell-shaped design, effectively attenuates high-frequency ambient noise and better isolates localized machine sounds \cite{kim2025control}. Leveraging the complementary nature of both sensors, we construct a dataset capturing both global and localized acoustic characteristics, supporting tasks from general classification to fine-grained anomaly detection. All recordings are made at 48,000 Hz, mono, 16-bit resolution to ensure high-fidelity signal acquisition.

\begin{figure}[t]
            \centering
            \includegraphics[width=1.0\linewidth]{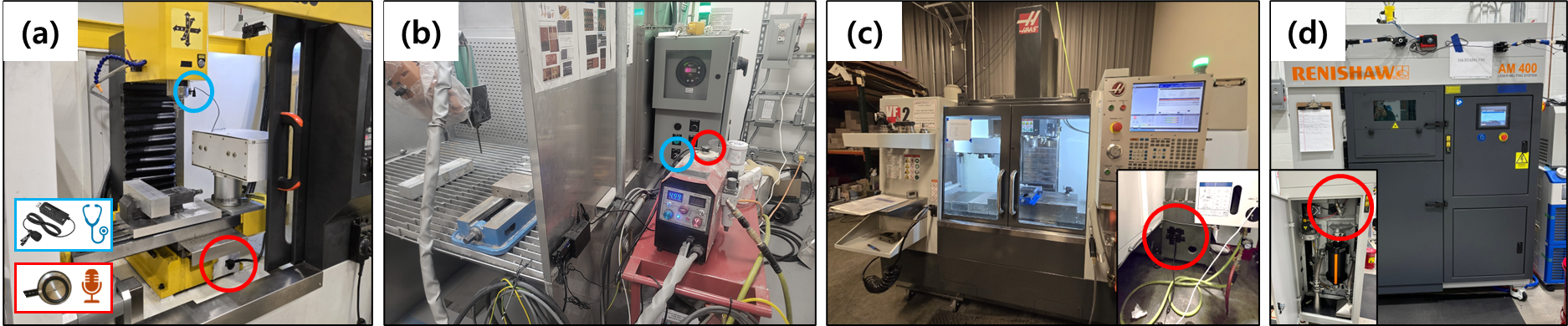}
            \caption{\textbf{Target Machines and Sound Sensor Placements.} Locations of microphone (blue circle) and stethoscope (red circle) sensors on various industrial machines,  (a) Yornew CNC machine (VMC-300): A microphone is positioned beside the spindle, and a stethoscope sensor is attached to the rigid body beneath the CNC table. (b) Cold spray powder feeder (BaltiCold Spray LTD, CSM 108.2): A stethoscope sensor is attached to the feeder, with an adjacent microphone. (c) Haas CNC machine (VF-2): A stethoscope sensor is placed on the lower right base. (d) Renishaw AM machine (AM-400): A stethoscope sensor is installed near the working area, inside the left powder handling panel.}
            \label{fig:machine2}
            \end{figure}    
            
\subsection{Data Source and Distribution}
Figure \ref{fig:machine2} illustrates the data acquisition process and sensor mounting positions. To ensure accurate acquisition of vibration-coupled signals, the stethoscope sensors are mounted on rigid machine structures such as motor bases or mainframes. This strategy avoids damping and distortion often caused by attaching sensors to flexible components like sheet-metal panels and allows for capturing consistent and representative machine acoustic signatures.

The dataset (see Table \ref{tab:tab1}) covers diverse manufacturing processes, materials, and operating conditions, enabling the representation of industrial acoustic features. It includes recordings from CNC cutting operations, Additive Manufacturing (AM) processes, and designed anomaly scenarios. For cutting, data are collected from two CNC machines: Haas VF-2 and Yornew VMC-300. VF-2 recordings include inactivity, machining, and warm-up sounds. VMC-300 machines aluminum (Al-6060) under varying feed rates and spindle speeds to induce chatter—a self-excited vibration that degrades surface finish and tool life by exciting the system’s natural frequencies. Additional metal processing sounds are collected from an APEC SK2540 CNC machine without annotation. For AM processes, the dataset includes recordings from Renishaw’s LPBF and FormAlloy’s DED systems, capturing both idle and operational states. Events such as fan activation, axis motion, and laser operation are reflected in the acoustic signal. Cold spray data are collected using a stethoscope sensor mounted on the powder feeder and a microphone placed in an open area. This setup enabled detection of anomalies such as gas flow loss, powder clogging, and depletion. Lastly, a microphone installed in a multi-machine shop floor captured ambient industrial noise including machine operations, fan rotation, and high-pressure air.

\begin{table}[!h]
  \caption{\textbf{DINOS Dataset.}  DINOS comprises 74,149 samples totaling over 1,093 hours, providing a comprehensive reference for developing and benchmarking diagnostic and monitoring systems across diverse industrial acoustic environments.} 
  \label{tab:tab1}
  \centering
  \resizebox{\textwidth}{!}{%
  \begin{tabular}{lcccl|lcccl}
    \toprule    
    \textbf{Category}   & \textbf{Sensor type}   & \textbf{Samples}   & \textbf{Duration}   & \textbf{Distribution (Samples)}  & \textbf{Category}   & \textbf{Sensor type}   & \textbf{Samples}   & \textbf{Duration}   & \textbf{Distribution (Samples)} \\
    \midrule
    CNC (SK2540)            & Stethoscope   & 21,570    & 59 s   & Pretraining (1,851)  &   AM-LPBF (RenishawL)     & Stethoscope   & 524       & 1 s    & Benchmarking (524)  \\
    AM-LPBF (RenishawR)     & Stethoscope   & 21,600    & 59 s   & Pretraining (1,851)  &   CNC (VMC-300)           & Stethoscope   & 461       & 1 s     & Benchmarking (461)  \\        
    AM-DED (FormAlloy)      & Stethoscope   & 21,600    & 59 s   & Pretraining (1,851)  &   CNC (VMC-300)           & Microphone    & 461       & 1 s     & Benchmarking (461)  \\
    Shop floor              & Microphone    & 1,851     & 59 s   & Pretraining (1,851)  &   AM-ColdSpray            & Stethoscope   & 2,455     & 1 s    & Benchmarking (2,455)  \\        
    CNC (VF-2)              & Stethoscope   & 1,118     & 1 s     & Benchmarking (1,118) &  AM-ColdSpray            & Microphone    & 2,509     & 1 s    & Benchmarking (2,509)  \\ 
    \bottomrule
  \end{tabular}
  }
\end{table}

\section{IMPACT Model}
\subsection{Pretraining Datasets}
To pretrain IMPACT, an equal number of samples is selected from four categories in DINOS to avoid learning bias. Data collected using stethoscope sensors are used to capture localized machine-specific acoustic characteristics, while shop floor recordings from microphones are included to reflect blended and dynamic industrial sounds. Specifically, 1,851 samples per category are selected, each lasting 59 seconds, totaling 121 hours of audio (see Table~\ref{tab:tab1}). Prior to pretraining, all samples are normalized using RMS and Z-score normalization and then segmented into 1-second clips while maintaining the sampling rate, resolution, and bit depth.
\subsection{IMPACT Architecture}
\begin{figure}[ht]
\centering
\includegraphics[width=1.0\linewidth]{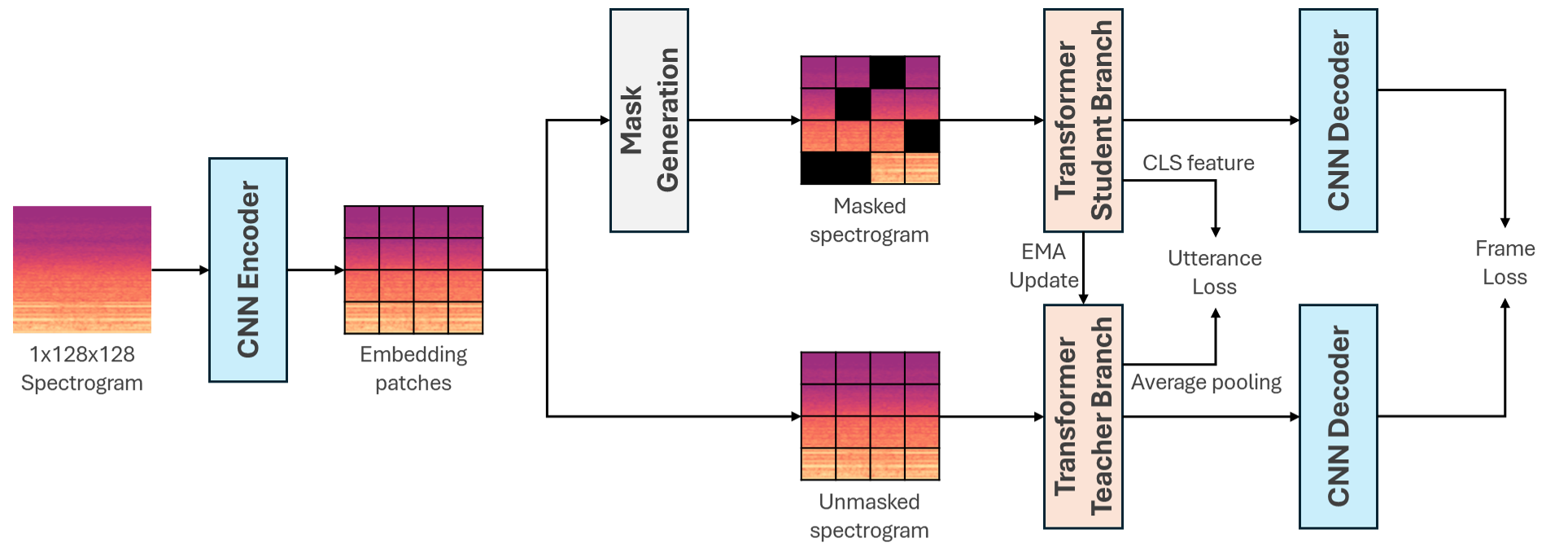}
\caption{\textbf{Architecture of IMPACT.} Overall pipeline with student-teacher branches.The student is trained using a reconstruction and alignment objective, while the teacher is updated via EMA. The detailed hyperparameters are presented in Appendix A.3.}
\label{fig:IMPACT_Arch}
\end{figure}

As shown in Figure \ref{fig:IMPACT_Arch}, IMPACT adopts the Efficient Audio Transformer (EAT) architecture \cite{chen2024eat}, comprising two identical sub-models: a student and a teacher. Each 1-second machine sound clip is transformed into a log-Mel spectrogram of dimension $1 \times 128 \times 128$ using the following parameters: 2,048 FFT (Fast Fourier Transform) Points; 2,048 Window Length; 376 Hop Length; 128 Mel Bands; 80 Top Decibels. These spectrograms are first processed by a CNN encoder, then split into $16 \times 16$ patches and embedded with fixed positional encoding. The student model receives masked spectrograms with a masking ratio of 0.7, while the teacher processes the unmasked input. One of the teacher’s outputs is obtained via average pooling across Transformer layers and compared to the student’s CLS token using mean squared error, forming the utterance-level loss. A CNN decoder is applied at the end of both models for reconstruction, with the frame-level loss computed using the Huber loss. IMPACT jointly optimizes both objectives as follows: $\mathcal{L}_{\text{total}} = \mathcal{L}_{f} + \lambda \mathcal{L}_{u}$, where $\mathcal{L}_f$ is the frame-level loss, $\mathcal{L}_u$ is the utterance-level loss, and $\lambda = 0.1$ controls their relative weight. This dual-objective strategy allows IMPACT to learn both fine-grained local details and global temporal structure. The teacher is updated using an Exponential Moving Average (EMA) of the student’s weights after every training epoch, with gradients blocked from flowing into the teacher during a training epoch. All preprocessing, training, and postprocessing were conducted using PyTorch 2.7.0 on an Ubuntu 22.04.5 LTS system, equipped with an AMD Ryzen Threadripper Pro 7975WX, 128 GB RAM, and NVIDIA RTX A6000 Ada. 

\section{Benchmarking}
\subsection{Benchmark Datasets}
Table \ref{tab:tab4} outlines the benchmark datasets, comprising samples from four machines, spanning 30 distinct downstream tasks. None of the benchmark samples are included in the pretraining data.

RenishawL and RenishawR share the same type but distinct machines. RenishawL's sound dataset is used for detecting machine on/off status. The data of VF2, a CNC machine, is used to monitor both machining activity and warm-up cycle operation. Yornew's data, another CNC machine, is used to classify machining states across 17 modes, which vary in Cutting Depth (CD), Material Removal Rate (MRR), spindle speed (Revolutions Per Minute, RPM), and chatter existence. Data for these three machines are collected using a stethoscope sensor. Cold spray is a metal additive manufacturing system driven by high-pressure gas. It is used to classify four operational states—normal, depleted powder, powder clogging, and no gas supply—based on sound recorded from both a microphone and a stethoscope sensor (total 8 downstreams). Cold spray is substantially different from the machines used in pretraining, making it a valuable benchmark for evaluating the model’s scalability across machine types and sensor modalities.

For benchmarking, 20\% of the samples from each downstream task are taken to train a single fully connected layer with 256 hidden units, initialized from scratch, on top of representations extracted from the frozen baseline and IMPACT encoders. The remaining 80\% are dedicated to evaluation. This process is repeated ten times with different splits, and the mean and standard deviation across these runs are reported for each downstream task and the overall model performance per machine.

\begin{table}[h!]
    \caption{\textbf{Benchmarking Dataset.} The dataset comprises 30 downstream tasks covering binary and multi-class classification across four machine types. It is designed to evaluate robustness to operational variation, sensor modality, and domain shift. (T1-T22: Stethoscope-sourced data)}
    \label{tab:tab4}
    \centering
    \resizebox{\textwidth}{!}{%
    \begin{tabular}{llll|llll}
        \toprule
        \textbf{Machine} & \textbf{ID} &\textbf{Description / Parameter} &  \textbf{Samples} & \textbf{Machine} & \textbf{ID} &\textbf{Description / Parameter} &  \textbf{Samples}\\
        \midrule
        RenishawL & T1 & On & 386 & Yornew & T16 & CD: 3.0, MRR: 76.2, RPM: 12K, Chatter: Y & 26\\
        RenishawL & T2 & Off & 138 & Yornew & T17 & CD: 3.0, MRR: 76.2, RPM: 11K, Chatter: N & 26 \\
        VF2 & T3 & On & 260 & Yornew & T18 & CD: 3.0, MRR: 76.2, RPM: 11K, Chatter: Y & 24 \\
        VF2 & T4 & Off & 504 & Yornew & T19 & CD: 3.0, MRR: 76.2, RPM: 10K, Chatter: Y & 26 \\
        VF2 & T5 & Warm-up & 354 & Yornew & T20 & CD: 3.0, MRR: 76.2, RPM: 9K, Chatter: Y & 26 \\
        Yornew & T6  & CD: 0.3, MRR: 7.62, RPM: 8K, Chatter: N & 78 & Yornew & T21 & CD: 3.0, MRR: 76.2, RPM: 8K, Chatter: Y & 24 \\
        Yornew & T7  & CD: 0.3, MRR: 7.62, RPM: 12K, Chatter: N & 78 & Yornew & T22 & CD: 3.0, MRR: 76.2, RPM: 7K, Chatter: Y & 26 \\
        Yornew & T8  & CD: 0.3, MRR: 7.62, RPM: 4K, Chatter: N & 78 & ColdSpray & T23 & Normal (stethoscope) & 1,214 \\
        Yornew & T9  & CD: 0.5, MRR: 12.7, RPM: 12K, Chatter: N & 74 & ColdSpray & T24 & Depleted powder (stethoscope) & 472 \\
        Yornew & T10 & CD: 0.5, MRR: 12.7, RPM: 8K, Chatter: N & 76 & ColdSpray & T25 & Powder clogging (stethoscope) & 385 \\
        Yornew & T11 & CD: 0.5, MRR: 12.7, RPM: 4K, Chatter: N & 78 & ColdSpray & T26 & No gas supply (stethoscope) & 384 \\
        Yornew & T12 & CD: 1.0, MRR: 25.4, RPM: 4K, Chatter: Y & 78 & ColdSpray & T27 & Normal (microphone) & 1,268 \\
        Yornew & T13 & CD: 1.0, MRR: 25.4, RPM: 8K, Chatter: N & 78 & ColdSpray & T28 & Depleted powder (microphone) & 472 \\
        Yornew & T14 & CD: 1.0, MRR: 25.4, RPM: 12K, Chatter: Y & 74 & ColdSpray & T29 & Powder clogging (microphone) & 385 \\
        Yornew & T15 & CD: 3.0, MRR: 76.2, RPM: 12K, Chatter: N & 52 & ColdSpray & T30 & No gas supply (microphone) & 384 \\
    \bottomrule
  \end{tabular}
  }
\end{table}

\subsection{Benchmark Baselines \& Metrics}
Along with IMPACT, we benchmark four widely used pretrained sound models, one fine-tunedd sound model, and two domain-specific models. \textbf{OpenSMILE} \cite{eyben2010opensmile} is adopted a feature extraction toolkit for speech and audio analysis. We use the ComParE feature set, a standardized configuration commonly employed in paralinguistic and affective computing tasks \cite{schuller2016interspeech}. \textbf{CLAP}~\cite{elizalde2023clap} (Contrastive Language-Audio Pretraining) is a multimodal model trained on paired audio-text data across a large and diverse set of sources, enabling zero-shot audio classification and flexible representation learning. \textbf{VGGish} \cite{hershey2017cnn} is a CNN-based model derived from the VGG architecture and trained on YouTube audio via the AudioSet dataset. It provides general-purpose embeddings widely used in audio classification tasks. \textbf{AudioMAE} \cite{huang2022masked} is a transformer-based self-supervised model trained using masked autoencoding. It is pretrained on AudioSet, ESC-50, Speech Commands, and VoxCeleb, and has shown strong performance in a variety of downstream audio tasks. To assess the utility of the DINOS dataset, we fine-tune AudioMAE using the subset of DINOS and compare its performance to the original pretrained version. For domain-specific comparison, we include \textbf{OPERA} \cite{zhang2024towards}, a family of foundation models originally designed for respiratory acoustic sensing. Among the three available variants, we evaluate two transformer-based models: OPERA-CT and OPERA-GT trained with contrastive and generative learning objectives, respectively. Both models are designed to process respiratory sounds via a stethoscope, which aligns with the data acquisition methodology used in DINOS, making them suitable baselines for evaluating domain adaptation.

We employ the standard linear probe protocol for downstream task evaluation \cite{zhang2024towards}, where a single fully connected layer with 256 hidden units is trained on top of representations extracted from frozen baseline and IMPACT encoders. To mitigate class imbalance, we report classification performance using \textbf{F1 score} for all models. Each model receives input data processed with the same preprocessing steps: RMS and Z-score normalization. Using 1-second clips, padding is applied if a model requires a longer input sequence. For baseline models, we follow their original implementation procedures.

\subsection{Benchmark Results}

\begin{table}[h!]
  \caption{\textbf{Overall Model Performance (F1 Score ± Standard deviation) per Machine.} Each entry shows the mean F1 score for each machine and its standard deviation across multiple runs. Higher F1 scores indicate better classification performance.}
  \label{tab:f1_per_machine}
  \centering
  \resizebox{\textwidth}{!}{%
  \begin{tabular}{lcccccccc}
    \toprule
    \textbf{Machine}
      & \textbf{OpenSMILE}        & \textbf{CLAP}
      & \textbf{VGGish}           & \textbf{OPERA-CT}
      & \textbf{OPERA-GT}         & \textbf{AudioMAE-PreT.}
      & \textbf{AudioMAE-FineT.}  & \textbf{IMPACT} \\
    \midrule
    RenishawL
      & 0.4239 ± 0.0000  & \textbf{1.0000 ± 0.0000}
      & 0.9857 ± 0.0130  & \textbf{1.0000 ± 0.0000}
      & 0.9408 ± 0.0546  & 0.9796 ± 0.0132
      & 0.9949 ± 0.0163  & \textbf{1.0000 ± 0.0000} \\
    VF2
      & 0.2536 ± 0.0679  & 0.9425 ± 0.0085
      & 0.8536 ± 0.0221  & 0.9382 ± 0.0120
      & 0.9319 ± 0.0084  & 0.9292 ± 0.0097
      & 0.9568 ± 0.0060  & \textbf{0.9618 ± 0.0073} \\
    Yornew
      & 0.0169 ± 0.0054  & 0.8549 ± 0.0307
      & 0.5844 ± 0.0288  & 0.8193 ± 0.0200
      & 0.8445 ± 0.0316  & 0.7733 ± 0.0331
      & 0.8305 ± 0.0392  & \textbf{0.8915 ± 0.0318} \\
    ColdSpray
      & 0.0357 ± 0.0148  & 0.8288 ± 0.0089
      & 0.7603 ± 0.0111  & 0.9197 ± 0.0063
      & 0.8931 ± 0.0081  & 0.7598 ± 0.0093
      & 0.8778 ± 0.0081  & \textbf{0.9525 ± 0.0116} \\
    \bottomrule
  \end{tabular}
  }
\end{table}

\begin{table}[h!]
  \caption{\textbf{Per-Class Model Performance (F1 Score ± Standard deviation) for Downstream Tasks.} F1 scores and standard deviations are reported for all 30 tasks. Higher F1 scores reflect more accurate and reliable task-specific classification. (Bold text represents the highest performance)}
  \label{tab:f1_t1_t30}
  \centering
  \resizebox{\textwidth}{!}{%
  \begin{tabular}{lcccccccc}
    \toprule
    \textbf{Task}
      & \textbf{OpenSMILE} & \textbf{CLAP} & \textbf{VGGish}
      & \textbf{OPERA-CT} & \textbf{OPERA-GT}
      & \textbf{AudioMAE-PreT.} & \textbf{AudioMAE-FineT.}
      & \textbf{IMPACT} \\
    \midrule
    T1  & 0.8477 ± 0.0000  & \textbf{1.0000 ± 0.0000}
        & 0.9926 ± 0.0061  & \textbf{1.0000 ± 0.0000}
        & 0.9730 ± 0.0232  & 0.9896 ± 0.0060
        & 0.9975 ± 0.0076  & \textbf{1.0000 ± 0.0000} \\
    T2  & 0.0000 ± 0.0000  & \textbf{1.0000 ± 0.0000}
        & 0.9788 ± 0.0185  & \textbf{1.0000 ± 0.0000}
        & 0.9085 ± 0.0804  & 0.9695 ± 0.0191
        & 0.9922 ± 0.0233  & \textbf{1.0000 ± 0.0000} \\
    \midrule
    T3  & 0.1322 ± 0.1840  & 0.9640 ± 0.0078
        & 0.8635 ± 0.0278  & 0.9691 ± 0.0035
        & 0.9622 ± 0.0071  & 0.9197 ± 0.0135
        & 0.9668 ± 0.0080  & \textbf{0.9714 ± 0.0083} \\
    T4  & 0.6286 ± 0.0095  & 0.9358 ± 0.0076
        & 0.8573 ± 0.0233  & 0.9307 ± 0.0138
        & 0.9225 ± 0.0099  & 0.9261 ± 0.0097
        & 0.9550 ± 0.0071  & \textbf{0.9573 ± 0.0076} \\
    T5  & 0.0000 ± 0.0000  & 0.9277 ± 0.0110
        & 0.8400 ± 0.0240  & 0.9147 ± 0.0198
        & 0.9110 ± 0.0109  & 0.9417 ± 0.0104
        & 0.9486 ± 0.0095  & \textbf{0.9567 ± 0.0097} \\
    \midrule
    T6  & 0.0000 ± 0.0000  & 0.9389 ± 0.0142
        & 0.7164 ± 0.0586  & 0.9127 ± 0.0302
        & 0.9472 ± 0.0178  & 0.8503 ± 0.0739
        & 0.9248 ± 0.0236  & \textbf{0.9582 ± 0.0208} \\
    T7  & 0.0000 ± 0.0000  & 0.7516 ± 0.0463
        & 0.5455 ± 0.0705  & 0.6408 ± 0.0538
        & 0.6670 ± 0.0912  & 0.7101 ± 0.0850
        & 0.7225 ± 0.0751  & \textbf{0.7979 ± 0.0702} \\
    T8  & 0.0000 ± 0.0000  & 0.4778 ± 0.1144
        & 0.3799 ± 0.1199  & 0.4623 ± 0.0764
        & 0.3191 ± 0.0848  & 0.3487 ± 0.1942
        & 0.3643 ± 0.1410  & \textbf{0.5676 ± 0.0849} \\
    T9  & 0.0000 ± 0.0000  & \textbf{0.9239 ± 0.0924}
        & 0.4352 ± 0.0934  & 0.6569 ± 0.1027
        & 0.8275 ± 0.1852  & 0.8311 ± 0.0889
        & 0.7818 ± 0.1553  & 0.8419 ± 0.1429 \\
    T10 & 0.0000 ± 0.0000  & \textbf{0.8782 ± 0.1346}
        & 0.6342 ± 0.1360  & 0.6779 ± 0.1056
        & 0.8529 ± 0.1611  & 0.7497 ± 0.1582
        & 0.7636 ± 0.1888  & 0.8648 ± 0.1475 \\
    T11 & 0.0000 ± 0.0000  & 0.7956 ± 0.1344
        & 0.5498 ± 0.1108  & 0.8982 ± 0.0719
        & 0.9338 ± 0.1302  & \textbf{0.9483 ± 0.1330}
        & 0.8753 ± 0.1677  & 0.9475 ± 0.1339 \\
    T12 & 0.0000 ± 0.0000  & 0.8465 ± 0.0913
        & 0.6652 ± 0.1292  & 0.9678 ± 0.0808
        & 0.9795 ± 0.0484  & 0.9931 ± 0.0146
        & 0.9884 ± 0.0116  & \textbf{1.0000 ± 0.0000} \\
    T13 & 0.0000 ± 0.0000  & 0.8621 ± 0.1129
        & 0.4352 ± 0.1598  & 0.8995 ± 0.1177
        & 0.9389 ± 0.1247  & 0.8285 ± 0.1124
        & 0.8964 ± 0.1507  & \textbf{0.9504 ± 0.1057} \\
    T14 & 0.0000 ± 0.0000  & 0.9058 ± 0.1306
        & 0.6178 ± 0.1455  & 0.9166 ± 0.1519
        & 0.9304 ± 0.1274  & 0.8300 ± 0.1163
        & 0.8512 ± 0.1290  & \textbf{0.9365 ± 0.1197} \\
    T15 & 0.0947 ± 0.0833  & \textbf{0.8564 ± 0.0345}
        & 0.4109 ± 0.0442  & 0.8445 ± 0.0427
        & 0.7884 ± 0.0530  & 0.5615 ± 0.0612
        & 0.7673 ± 0.0645  & 0.8458 ± 0.0428 \\
    T16 & 0.0529 ± 0.0707  & 0.7271 ± 0.0464
        & 0.4269 ± 0.0592  & 0.7106 ± 0.0517
        & 0.6428 ± 0.0304  & 0.4153 ± 0.0801
        & 0.6343 ± 0.0799  & \textbf{0.7477 ± 0.0565} \\
    T17 & 0.0968 ± 0.0444  & 0.8577 ± 0.0277
        & 0.6598 ± 0.0512  & 0.7822 ± 0.0497
        & 0.8156 ± 0.0245  & 0.8085 ± 0.0354
        & 0.8093 ± 0.0318  & \textbf{0.8772 ± 0.0346} \\
    T18 & 0.0000 ± 0.0000  & 0.9248 ± 0.0166
        & 0.7044 ± 0.0482  & 0.9401 ± 0.0224
        & 0.9539 ± 0.0213  & 0.8517 ± 0.0576
        & 0.9189 ± 0.0208  & \textbf{0.9580 ± 0.0268} \\
    T19 & 0.0305 ± 0.0611  & 0.9503 ± 0.0162
        & 0.5479 ± 0.0535  & 0.8844 ± 0.0369
        & 0.9304 ± 0.0245  & 0.8043 ± 0.0671
        & 0.9464 ± 0.0439  & \textbf{0.9734 ± 0.0166} \\
    T20 & 0.0000 ± 0.0000  & 0.9435 ± 0.0247
        & 0.6907 ± 0.0256  & 0.8856 ± 0.0328
        & 0.9549 ± 0.0231  & 0.8525 ± 0.0526
        & 0.9683 ± 0.0227  & \textbf{0.9776 ± 0.0168} \\
    T21 & 0.0115 ± 0.0346  & 0.9402 ± 0.0104
        & 0.8211 ± 0.0297  & 0.9421 ± 0.0208
        & 0.9503 ± 0.0290  & 0.9248 ± 0.0273
        & \textbf{0.9665 ± 0.0137}  & 0.9651 ± 0.0168 \\
    T22 & 0.0000 ± 0.0000  & \textbf{0.9536 ± 0.0236}
        & 0.6936 ± 0.0358  & 0.9051 ± 0.0344
        & 0.9244 ± 0.0282  & 0.8381 ± 0.0537
        & 0.9386 ± 0.0327  & 0.9465 ± 0.0240 \\
    \midrule
    T23 & 0.0000 ± 0.0000  & 0.9381 ± 0.0081
        & 0.8540 ± 0.0103  & 0.9592 ± 0.0061
        & 0.9385 ± 0.0068  & 0.8762 ± 0.0109
        & 0.9556 ± 0.0056  & \textbf{0.9870 ± 0.0037} \\
    T24 & 0.0000 ± 0.0000  & 0.9506 ± 0.0097
        & 0.8928 ± 0.0104  & 0.9608 ± 0.0061
        & 0.9452 ± 0.0082  & 0.7750 ± 0.0187
        & 0.9454 ± 0.0111  & \textbf{0.9767 ± 0.0115} \\
    T25 & 0.0000 ± 0.0000  & 0.8702 ± 0.0218
        & 0.7699 ± 0.0215  & 0.9110 ± 0.0201
        & 0.8880 ± 0.0160  & 0.8289 ± 0.0169
        & 0.9038 ± 0.0141  & \textbf{0.9894 ± 0.0082} \\
    T26 & 0.0000 ± 0.0000  & 0.9500 ± 0.0102
        & 0.8323 ± 0.0249  & 0.9727 ± 0.0071
        & 0.9742 ± 0.0085  & 0.9877 ± 0.0071
        & 0.9974 ± 0.0041  & \textbf{0.9976 ± 0.0022} \\
    T27 & 0.1629 ± 0.1991  & 0.8700 ± 0.0104
        & 0.7950 ± 0.0089  & 0.9346 ± 0.0057
        & 0.9054 ± 0.0067  & 0.8018 ± 0.0089
        & 0.8975 ± 0.0139  & \textbf{0.9614 ± 0.0191} \\
    T28 & 0.1225 ± 0.1131  & 0.6127 ± 0.0193
        & 0.4743 ± 0.0267  & 0.8171 ± 0.0158
        & 0.7433 ± 0.0139  & 0.3964 ± 0.0365
        & 0.6807 ± 0.0416  & \textbf{0.8546 ± 0.0275} \\
    T29 & 0.0000 ± 0.0000  & 0.4386 ± 0.0352
        & 0.4958 ± 0.0585  & 0.8025 ± 0.0165
        & 0.7504 ± 0.0248  & 0.4185 ± 0.0270
        & 0.6476 ± 0.0250  & \textbf{0.8532 ± 0.0408} \\
    T30 & 0.0000 ± 0.0000  & 0.9998 ± 0.0005
        & 0.9683 ± 0.0145  & 0.9997 ± 0.0006
        & 0.9995 ± 0.0007  & 0.9942 ± 0.0023
        & 0.9940 ± 0.0050  & \textbf{1.0000 ± 0.0000} \\
    \bottomrule
  \end{tabular}
  }
\end{table}

To evaluate the sound models, we conduct benchmark experiments across 30 downstream tasks involving four industrial machines: RenishawL, VF2, Yornew, and Cold spray. All models are evaluated using the F1 score under a standardized linear probing protocol. The performance results are summarized in Table~\ref{tab:f1_per_machine} and Table~\ref{tab:f1_t1_t30}.
\paragraph{RenishawL (T1–T2): On/Off Classification of LPBF System.}
This experiment is selected as a minimum performance threshold for each model. This binary task requires recognizing subtle differences between the idle and active acoustic states of an LPBF machine. All high-capacity pretrained models achieve near-perfect scores, with IMPACT, CLAP, and OPERA-CT reaching 1.0000. In contrast, general-purpose models like OpenSMILE performs poorly (F1 < 0.43), reflecting their inability to resolve stable harmonics, operational periodicity, and structured broadband noise typical of industrial machine sounds.
\paragraph{VF2 (T3–T5): CNC Operation Mode Classification.}
The tasks involve identifying whether a CNC machine is in an active, idle, or warm-up state. These classification problems rely on subtle but periodic acoustic patterns, such as spindle rotations and axis movement pulses. In addition, distinguishing between tool-workpiece direct contact sounds during machining is essential. While CLAP, OPERA-CT, OPERA-GT, and pretrained AudioMAE demonstrate solid performance (F1 > 0.92), IMPACT consistently achieves higher scores (F1 > 0.96). This improvement can be attributed to IMPACT’s self-supervised pretraining on structurally repetitive machine audio, which allows it to capture temporal regularities that general-purpose, speech-trained models fail to learn. The performance gains observed when fine-tuning AudioMAE (F1 > 0.95) on industrial data further emphasize the importance of data that reflect the unique acoustic characteristics of manufacturing environments. Although OPERA models are also trained on stethoscope-recorded signals, their specialization in respiratory sounds which lack the mechanical complexity of industrial acoustics likely limits their generalization in this context.
\paragraph{Yornew (T6–T22): Multi-Class Machining State Recognition.}
This task group assesses the model's ability to distinguish among 17 distinct CNC machining configurations—varying in cutting depth, MRR, spindle speed (RPM), and chatter (vibration-induced defect during cutting processes) using a limited number of samples and multi-modal sounds recorded by both stethoscope and microphone sensors. Many tasks involve overlapping frequency bands and subtle transient variations, making classification particularly challenging. AudioMAE-fine-tunedd, trained on DINOS, shows improved performance in this group and demonstrates the benefit of exposure to stethoscope-recorded machine sounds. OPERA-CT and OPERA-GT also outperform general-purpose pretrained models, likely due to their training on stethoscopic acoustic data. Notably, OPERA-GT surpasses OPERA-CT in this group, likely due to the generative model's superior ability to capture comprehensive features arising from the diverse conditions in this task group \cite{zhang2024towards}. CLAP yields the best performance among the pretrained models due to its learned features from over 630 K samples with paired texts. However, all pretrained models underperform (F1 < 0.86) due to their limited sensitivity to fine-grained tonal and vibrational patterns. IMPACT achieves the best performance (F1 > 0.89), benefiting from its hybrid pretraining strategy that combines contrastive and generative learning across multi-modal inputs, effectively capturing both global semantic information and fine-grained temporal structures in industrial machine acoustics.
\paragraph{Coldspray (T23–T30): Fault Detection Across Sensor Types.}
These tasks evaluate the model’s ability to generalize to an unseen domain. Although Coldspray data is not included in pretraining, the model must detect abnormal gas flow and powder-related conditions. IMPACT achieves the highest performance across all scenarios and sensor types (F1 > 0.96), demonstrating strong robustness to both domain and modality shifts. Interestingly, while OPERA-CT and OPERA-GT perform well on stethoscope recordings (T23–T26), their overall performance degrades notably on microphone-based recordings (T27–T30). This suggests a degree of modality-specific overfitting and an inability to effectively handle background noise blending. Similarly, VGGish and AudioMAE-PreTrained also struggle with microphone data, possibly due to their nature, which does not require precise resolution. In contrast, IMPACT’s pretraining on both localized and blended acoustic data allows it to handle complex mixtures of machine sound and environmental noise. The effectiveness of this strategy is further supported by the substantial performance gain (from 0.76 to 0.88) observed in AudioMAE-fine-tunedd after exposure to the DINOS dataset.
\paragraph{Summary.}
OpenSMILE consistently shows the lowest performance across all task categories, with F1 scores approaching zero in multi-class and fault detection scenarios. While CLAP outperforms on several tasks, its reliance on large sound datasets with paired texts demands significant preparation resources. VGGish performs moderately on binary tasks but struggles with fine-grained classification and generalization to unseen domains, likely due to its limited temporal resolution and convolutional architecture. AudioMAE-PreTrained, despite its transformer-based structure, lacks industrial audio exposure, leading to suboptimal representations for machine perception. Fine-tuning AudioMAE on DINOS improves its performance across tasks, highlighting the impact of domain-specific data, though it still underperforms IMPACT, especially with limited samples. OPERA-CT and OPERA-GT, though trained on stethoscope-collected acoustic signals, generalize well to similar sensor conditions but degrade with microphone data, suggesting modality-specific overfitting and a narrower representation space tuned to respiratory rather than mechanical dynamics. In contrast, IMPACT outperforms all baseline models on 24 out of 30 downstream tasks. Its hybrid training objective—capturing global and fine-grained representations—combined with exposure to both localized and ambient acoustic signals from DINOS, enables robust temporal modeling and cross-modality generalization. These results compellingly show that general-purpose sound models, typically trained on human-centric or environmental sounds, fail to generalize to the structured, repetitive, and low-SNR characteristics of machine acoustics. These findings reinforce the necessity of domain-specialized foundation models for scalable, transferable industrial machine perception.

\section{Conclusion, Limitations, and Future Work}
In this work, we present IMPACT, the first foundation model specifically designed for industrial acoustic perception. IMPACT outperforms existing general-purpose and domain-specific sound models across a wide range of machine classification and fault detection tasks. Its superior performance is attributed to two key design decisions. 

First, the model is pretrained on DINOS, a large-scale dataset that captures distinct characteristics of real-world machine sounds. The importance of domain-specific data is highlighted by the significant performance gains observed in AudioMAE after fine-tuning on DINOS. Second, IMPACT leverages a hybrid pretraining objective that integrates both contrastive learning and generative reconstruction. This dual strategy enables the model to effectively capture both global and local acoustic features, combining the strengths observed separately in OPERA-CT and OPERA-GT models.

Despite its strong performance, IMPACT has several limitations. First, a subset of the Yornew machining tasks—particularly those involving subtle conditional variations in operations—remains challenging, indicating that further performance gains will require more extensive and balanced data collection across operational regimes. Second, although IMPACT is more lightweight than VGGish (62M), AudioMAE (86M), and both OPERA models (21–31M), it still consists of 18 million parameters \cite{zhang2024towards}. This makes it non-trivial to deploy on low-resource embedded devices in real-time industrial environments. Model compression and architecture distillation will be essential next steps toward enabling real-time inference in low-end edge computing scenarios.

Finally, we highlight a broader challenge in the industrial domain: data sharing remains highly constrained due to security and intellectual property concerns. This significantly limits the availability of high-quality, diverse datasets for training and evaluation. To address this bottleneck, we plan to explore synthetic data generation via physical models and digital twin systems. By augmenting limited real-world recordings with physics-informed synthetic data generation, we aim to amplify the dataset scale and diversity without compromising data confidentiality. We envision future versions of IMPACT to be trained on hybrid datasets that combine real and simulated audio, enabling further improvements in generalization and robustness under complex manufacturing conditions.
\newpage


\bibliography{neurips_2025}

\newpage
\appendix
\section{Technical Appendices and Supplementary Material}
\subsection{Machine Description}
\paragraph{AM-LPBF: AM400 (Renishaw).}
The AM400 is a laser powder bed fusion (LPBF) metal Additive Manufacturing (AM) system developed by Renishaw. LPBF is a metal 3D printing process that uses a laser to selectively melt metal powder in layers under an inert gas atmosphere. It supports materials such as Inconel 718 (RenishawL) and 316 stainless steel (RenishawR). The build chamber is maintained under inert argon gas to prevent oxidation. The official build volume is 250 × 250 × 300 mm.
\paragraph{AM-DED: L2 Series (FormAlloy).}
The L2 Series by FormAlloy is a directed energy deposition (DED) system designed for high-deposition-rate metal additive manufacturing. DED is a metal 3D printing method that melts and deposits material simultaneously using a laser and metal powder or wire. It processes materials such as A709 structural steel and 316H stainless steel using a coaxial laser and powder nozzle system. In this system, three powder feeders are used simultaneously to perform layered deposition. The laser power can reach up to 8 kW. It supports multi-material deposition, enabling the fabrication of complex, compositionally graded parts.
\paragraph{CNC: VMC300 (Yornew).}
The Yornew VMC-300 is a compact 5-axis CNC vertical milling center designed for prototyping and educational use. In this dataset, the machine is operated as a 3-axis CNC by detaching the A and C axes and is used to machine aluminum (Al-6060) material. A 1/4-inch diameter two‑flute end mill (YG‑1; 01047) is applied to the milling experiments. The machine provides a working volume of 300 × 150 × 100 mm, with a maximum spindle speed of 24,000 rpm. The maximum feed rate is 2,000 mm/min, and the spindle motor is rated at 750 W.
\paragraph{CNC: VF-2 (Haas).}
The Haas VF-2 is a high-performance 3-axis vertical machining center widely used in industrial and academic environments. For this study, a 1/2-inch two-flute end mill tool is utilized to machine ABS plastic. The VF-2 offers a working envelope of 762 × 406 × 508 mm, with a maximum spindle speed of 8,100 rpm and feed rates up to 16.5 m/min. The machine is equipped with an automatic tool changer (ATC) that supports up to 20 tools.
\paragraph{CNC: SK2540 (APEC CNC).}
Metal processing is performed on the real shop floor using the SK2540 model. This machine supports a working area of 4000 × 2500 × 1000 mm. It provides a rapid traverse rate of 60 m/min (XY) and 40 m/min (Z), and a 5 m/s² acceleration on all axes. The spindle operates at up to 24,000 rpm with a maximum spindle power of 75 kW, making it suitable for high-speed, high-precision aerospace aluminum machining tasks.
\paragraph{Coldspray: CSM 108.2 (BaltiCold Spray LTD).}
The CSM 108.2 is a cold spray system used for solid-state deposition of metallic powders. In the described experiment, copper (Cu) powders with a particle size range of 10–45~$\mu$m and a mean diameter (d50) of 17~$\mu$m are used. The powder is sprayed at room temperature using nitrogen gas at a constant gauge pressure of 0.7 MPa without preheating. The deposition process enables bonding without melting, preserving the original microstructure of the material. This system is commonly used for research on coating performance and defect detection under controlled conditions.

\subsection{Frequency Analysis}

\begin{figure}[h!]
            \centering
            \includegraphics[width=1.0\linewidth]{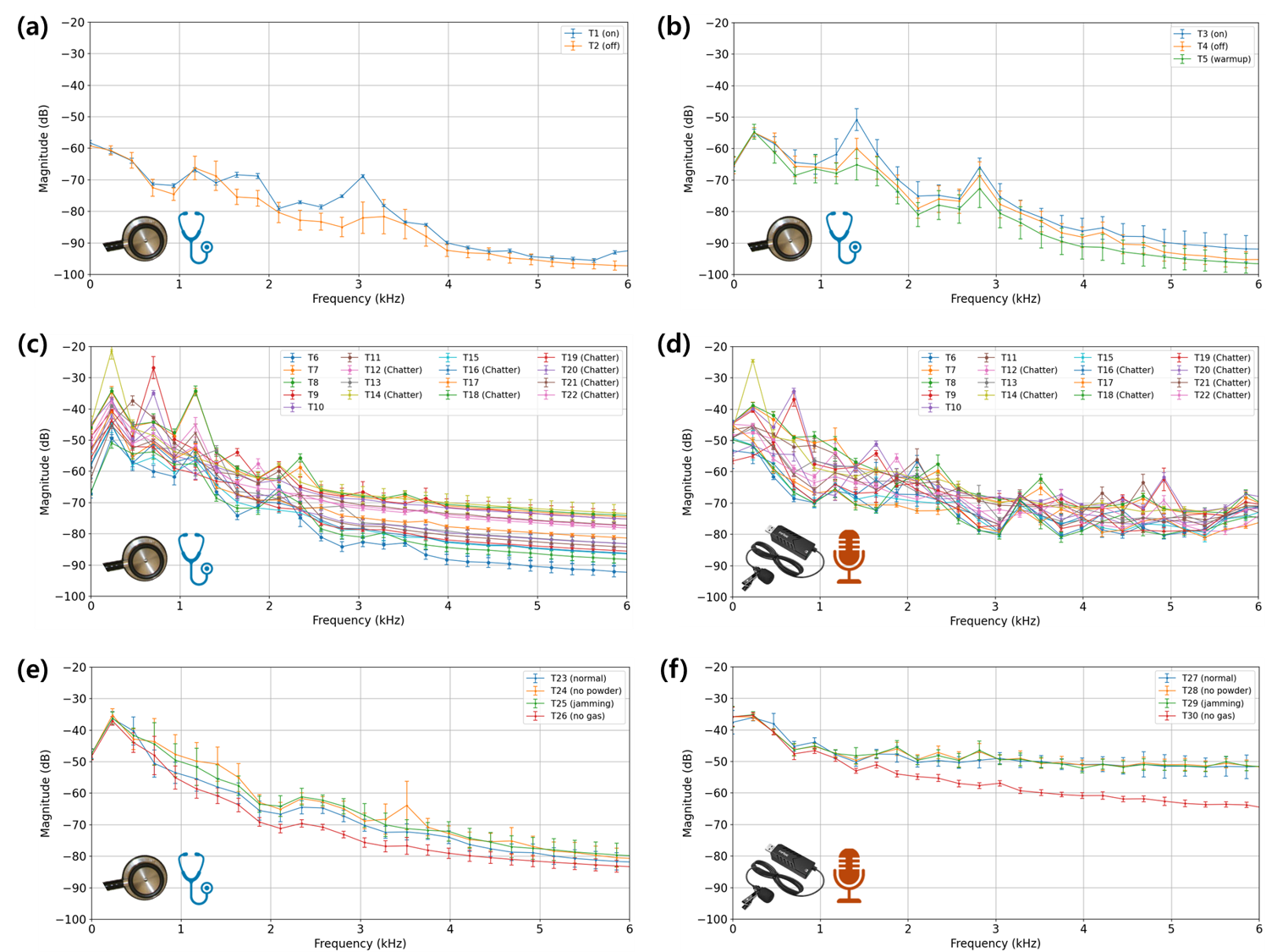}         
            \caption{\textbf{ Mean frequency spectra with standard deviation error bars for various machine operating states and sensor configurations.} (a) Frequency response of the Renishaw AM machine during operational and idle states, measured using a stethoscope acoustic sensor. (b) Spectra of the VF-2 CNC machine during operational, idle, and warm-up phases, also captured using a stethoscope. (c–d) Mean and variability of frequency spectra from the Yornew CNC machine at different cutting parameters and chatter conditions, recorded using (c) a stethoscope-type sensor and (d) a standard microphone. (e–f) Spectral characteristics of cold spray process anomalies—normal operation, powder absence, feed jamming, and gas cutoff—measured using (e) a stethoscope and (f) a microphone. All plots show the mean magnitude in decibels (dB) with standard deviation represented as error bars across frequency bins up to 6 kHz. }
            \label{fig:frequency}
            \end{figure}

Figure \ref{fig:frequency} presents the mean frequency spectra with standard deviation represented as error bars, obtained by applying FFT to every 2048-sample segment of the full acoustic recordings from each manufacturing process.

Figure \ref{fig:frequency}(a) illustrates the analysis of the Renishaw additive manufacturing process, where segments corresponding to active deposition are labeled as “on,” and all other segments as “off.” A clear spectral peak emerges around 3000 Hz, indicating the dominant operating frequency during the build phase.
Figure \ref{fig:frequency}(b) shows the frequency spectra of the VF-2 CNC machine under three distinct states: cutting, idle, and warm-up. Warm-up data are collected during non-cutting operations in which axis movements and spindle speed control are executed independently; the X-axis moved 304.8 mm, while the Y and Z axes moved 127 mm, alongside controlled acceleration of the spindle. Idle and non-cutting segments are categorized as “off,” which included background noise such as pump and tool changer operation. Cutting data are recorded during the machining of ABS workpieces.

Figures \ref{fig:frequency}(c) and \ref{fig:frequency}(d) display acoustic data collected from the Yornew CNC machine cutting aluminum, using a stethoscope sensor and a conventional microphone, respectively. These plots reveal both global and local signal features: globally, the overall magnitude varies according to MRR, while locally, distinct spectral peaks increase with spindle speed and relate to the number of tool flutes. This indicates that both broadband and frequency-specific information encode key process parameters. These multi-sensor measurements are used to assess IMPACT's ability to classify dynamic changes in process conditions based on both sensor domains.

Figures \ref{fig:frequency}(e) and \ref{fig:frequency}(f) focus on the Cold Spray process, where acoustic signals from the stethoscope and microphone sensors are analyzed to evaluate IMPACT’s capability to distinguish abnormal states independently. Compared to the normal case, the “no powder” condition produced high-frequency components near 1500 Hz and 3500 Hz, attributed to direct collisions between the vibration valve and powder feeder in the absence of powder. In the “jamming” state, insufficient damping from powder resulted in stronger structural vibration, leading to overall higher magnitude spectra. In contrast, the “no gas” condition—where powder ejection fails due to lack of compressed air—produced significantly lower spectral magnitudes.
The low-frequency amplification and high-frequency noise suppression characteristics of the stethoscope sensor are evident when comparing Figures \ref{fig:frequency}(c) to \ref{fig:frequency}(d) and Figures \ref{fig:frequency}(e) to \ref{fig:frequency}(f), respectively.

\subsection{IMPACT Architecture and Hyperparameters}
The IMPACT architecture, as detailed in Table~\ref{tab:impactparam}, is comprised of three main components: a CNN Encoder, a Transformer Encoder, and a CNN Decoder. The CNN Encoder consists of a single 2D convolutional layer (Conv2d) with Batch Normalization (BN) and GELU activation. It processes an input with 1 channel, transforming it into 32 channels using a $3 \times 3$ kernel, a stride of 2, and padding of 1. The Transformer Encoder is composed of 8 layers. Each layer operates with an embedding dimension of 384 and utilizes 16 attention heads. GELU is employed as the activation function within the transformer layers. The CNN Decoder begins with a Linear (Fully Connected, FC) layer that projects the input dimension from 384 to 512. This is followed by a series of four ConvTranspose2d layers, each incorporating Batch Normalization, GELU activation, and a 2x upsampling factor. These transpose convolutional layers consistently use a $4 \times 4$ kernel, a stride of 2, and padding of 1, while progressively changing channel dimensions from 128 to 128, then to 64, then to 32, and finally to 16. The decoder concludes with an output ConvTranspose2d layer that converts the 16 channels back to a single output channel, also using a $4 \times 4$ kernel, a stride of 2, and padding of 1. The model was trained for 10 epochs, with each epoch taking approximately 410 seconds.

\begin{table}[h!]
    \caption{\textbf{IMPACT Architecture and Hyperparameters.}}
    \label{tab:impactparam}
    \centering
    \resizebox{\linewidth}{!}{%
    \begin{tabular}{lccccl}
        \toprule
        \textbf{CNN Encoder} & Input Channels & Output Channels & Kernel & Stride & Padding \\
        \midrule
        Conv2d + BN + GELU & 1 & 32  & $3\times3$ & 2 & 1 \\
        \midrule
        \midrule
        \textbf{Transformer Encoder} & \#Layers & Dimension & \#Heads & Activation & — \\
        \midrule
        Transformer (Patch: 16x16) & 8       & 384      & 16    & GELU & — \\
        \midrule
        \midrule
        \textbf{CNN Decoder} & Input Channels & Output Channels & Kernel & Stride & Padding \\
        \midrule
        Linear (FC): $dimension\!=\!384\to512$ & —   & —  & —      & —       & — \\
        ConvTranspose2d + BN + GELU  & 128 & 128  & $4\times4$ & 2 & 1 \\
        ConvTranspose2d + BN + GELU  & 128 & 64   & $4\times4$ & 2  & 1 \\
        ConvTranspose2d + BN + GELU & 64  & 32   & $4\times4$ & 2 & 1 \\
        ConvTranspose2d + BN + GELU & 32  & 16   & $4\times4$ & 2 & 1 \\
        ConvTranspose2d (Output)             & 16  & 1    & $4\times4$ & 2 & 1 \\
        \bottomrule
        \end{tabular}%
        }
\end{table}

\subsection{Visualization of Benchmark Results}
\begin{figure}[h]
            \centering
            \includegraphics[width=1.0\linewidth]{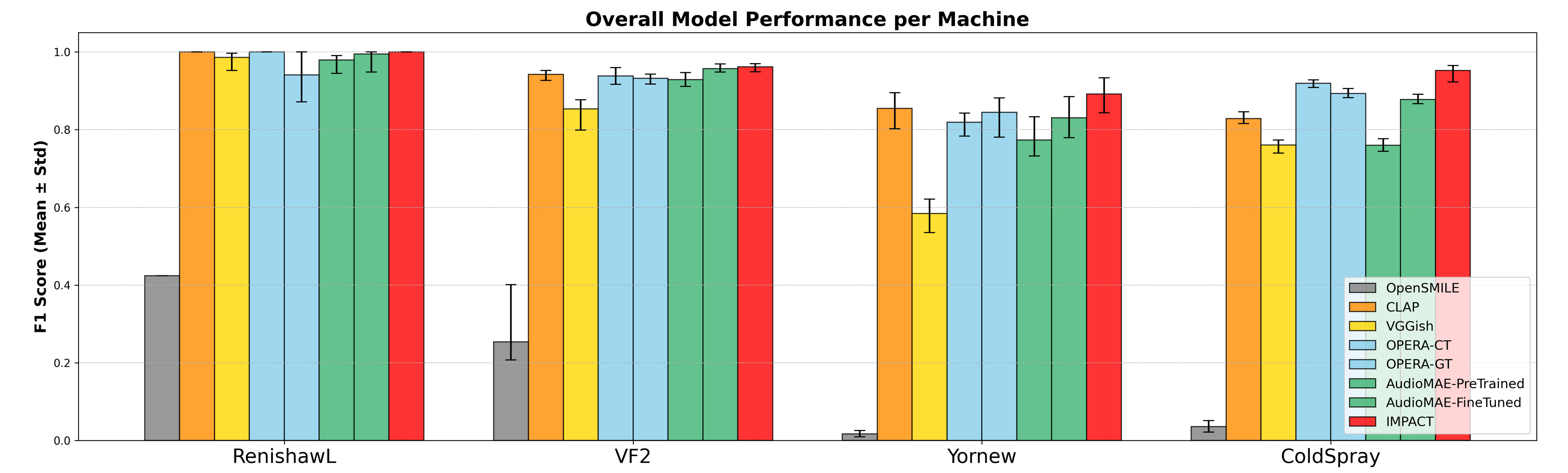}             
            \caption{\textbf{Graph of Overall Model Performance per Machine}. Bar chart comparing the average F1 scores (± standard deviation) of all baseline models per machine.}
            \label{fig:visualizedresults1}
            \end{figure}

\begin{figure}[h]
            \centering
            \includegraphics[width=1.0\linewidth]{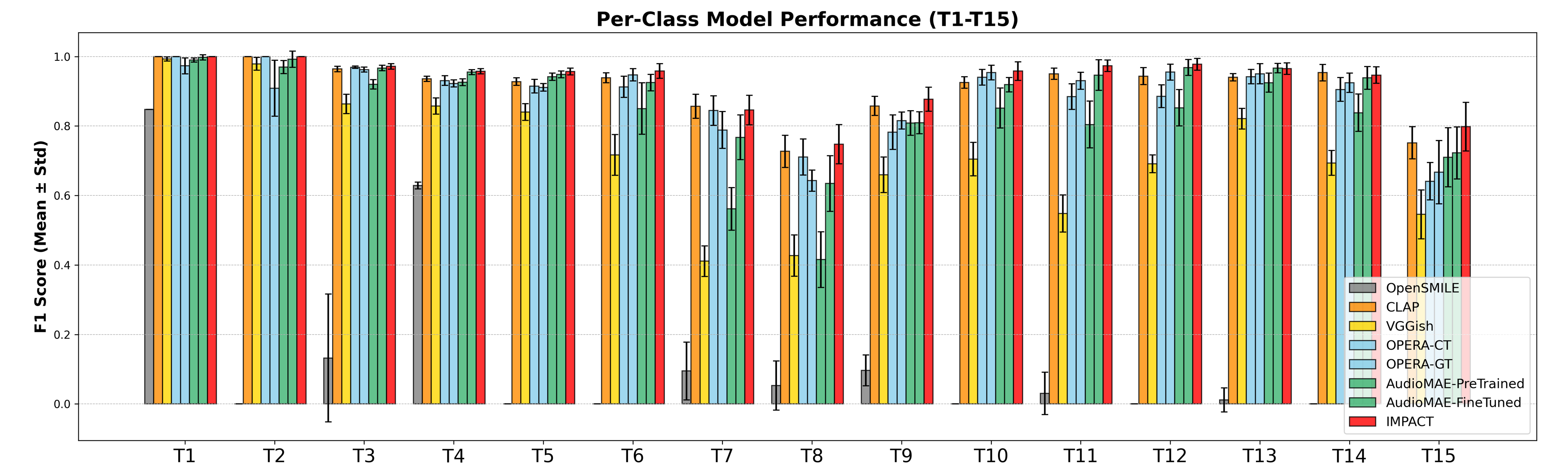}
            \includegraphics[width=1.0\linewidth]{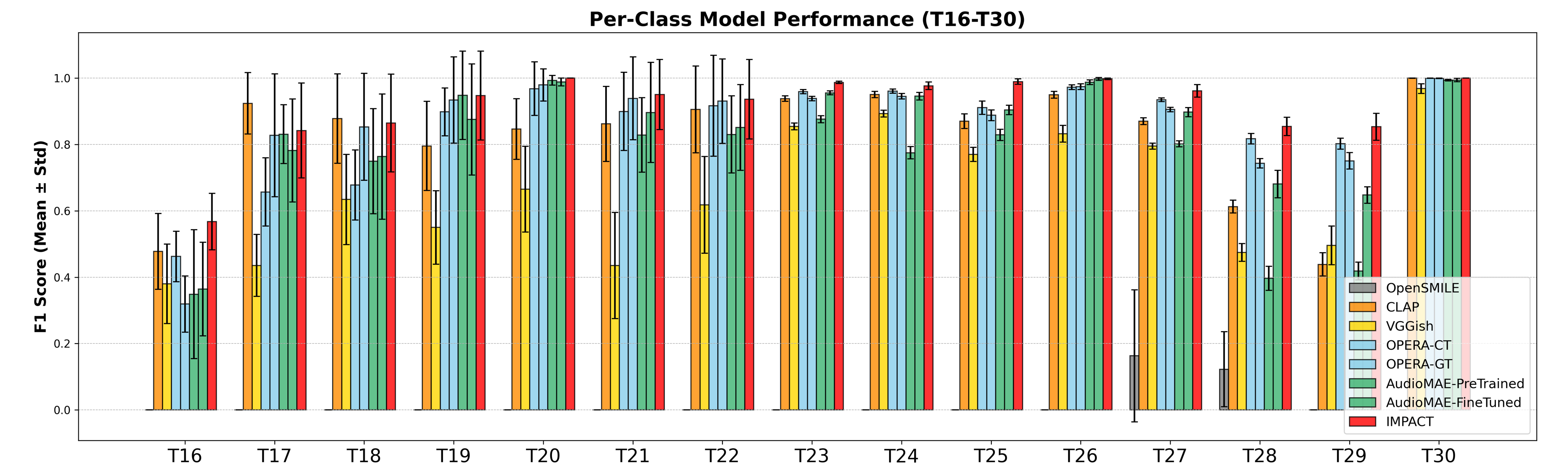}            
            \caption{\textbf{Graph of Per-Class Model Performance}. Bar chart comparing the average F1 scores (± standard deviation) of all baseline models for each downstream task.}
            \label{fig:visualizedresults2}
            \end{figure}

\begin{figure}[h]
            \centering
            \includegraphics[width=0.75\linewidth]{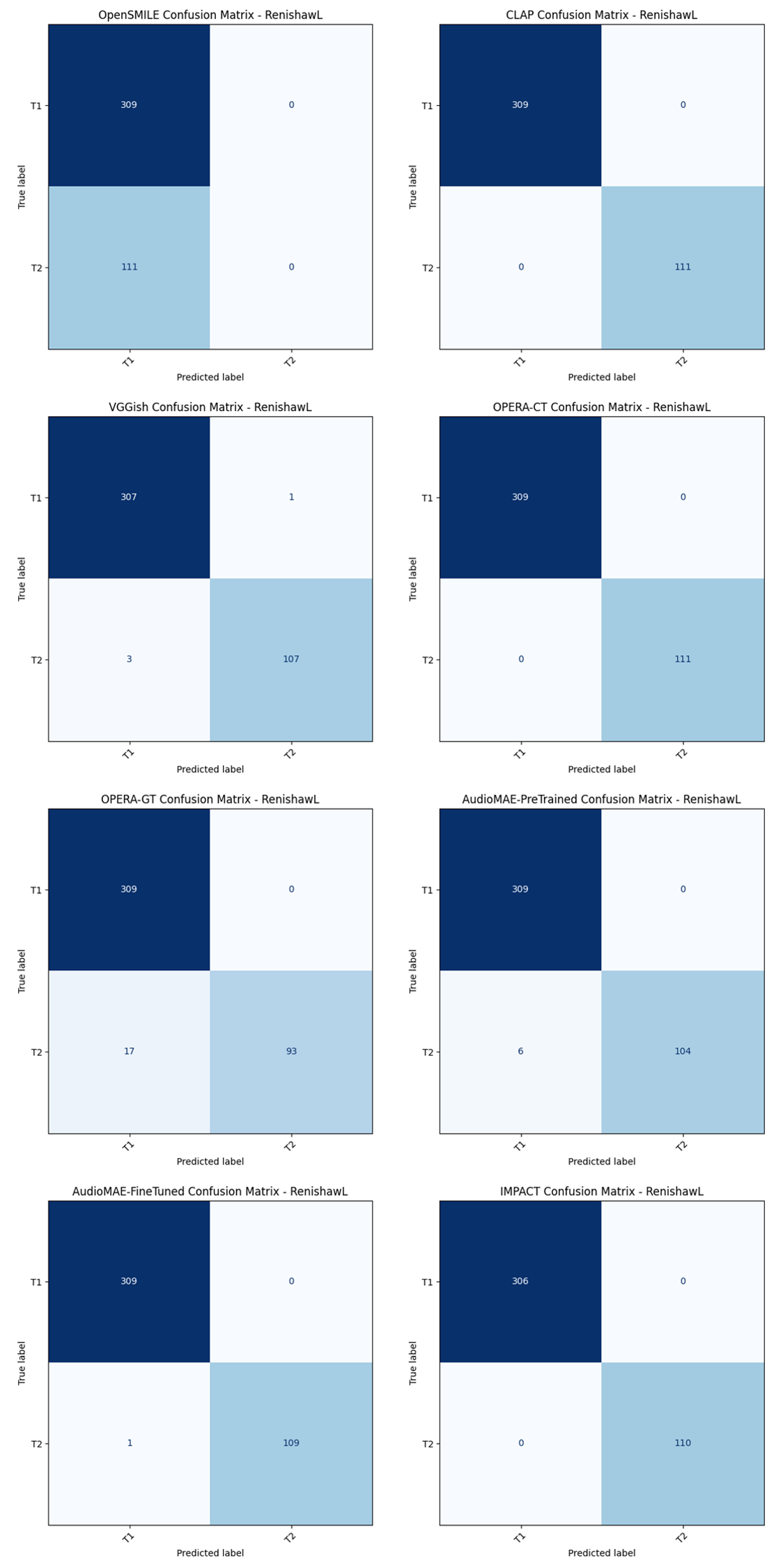}
            \caption{\textbf{Confusion Matrices for RenishawL}. Each cell in the table shows the count of instances predicted by the model for the corresponding class.}
            \label{fig:cm_renishawl}
            \end{figure}

\begin{figure}[h]
            \centering
            \includegraphics[width=0.75\linewidth]{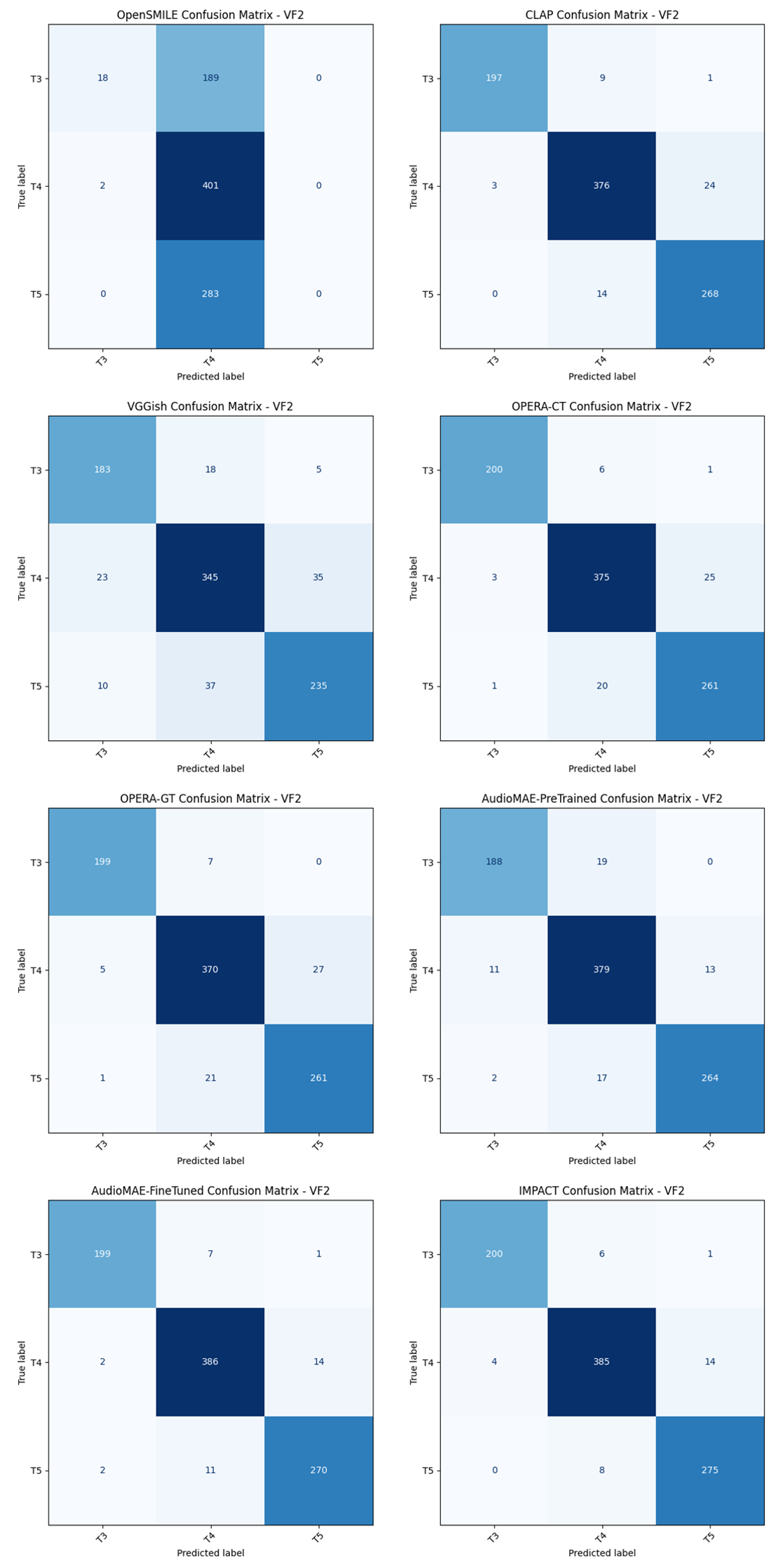}
            \caption{\textbf{Confusion Matrices for VF2}. Each cell in the table shows the count of instances predicted by the model for the corresponding class.}
            \label{fig:cm_vf2}
            \end{figure}

\begin{figure}[h]
            \centering
            \includegraphics[width=0.75\linewidth]{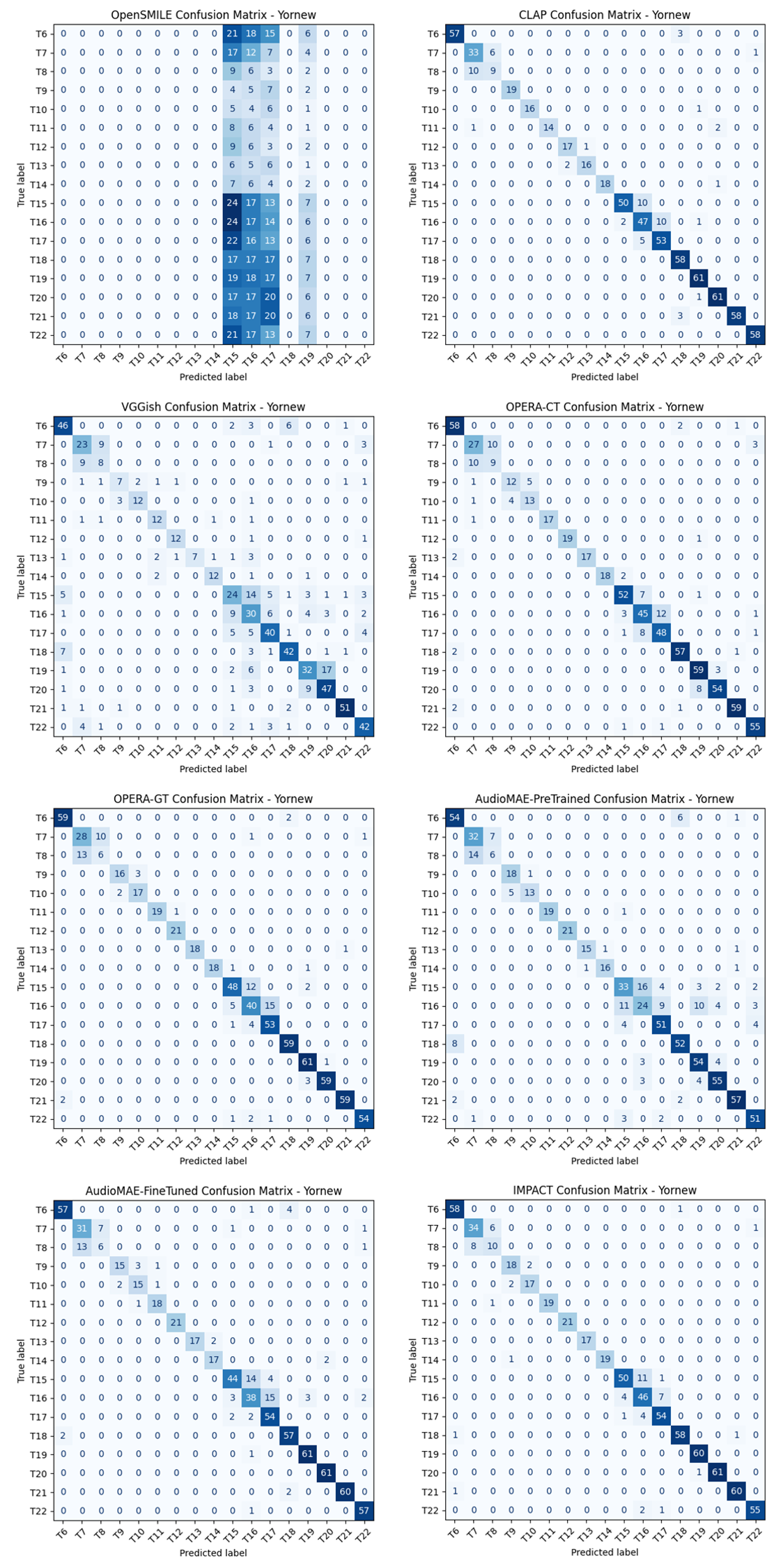}
            \caption{\textbf{Confusion Matrices for Yornew}. Each cell in the table shows the count of instances predicted by the model for the corresponding class.}
            \label{fig:cm_yornew}
            \end{figure}

\begin{figure}[h]
            \centering
            \includegraphics[width=0.75\linewidth]{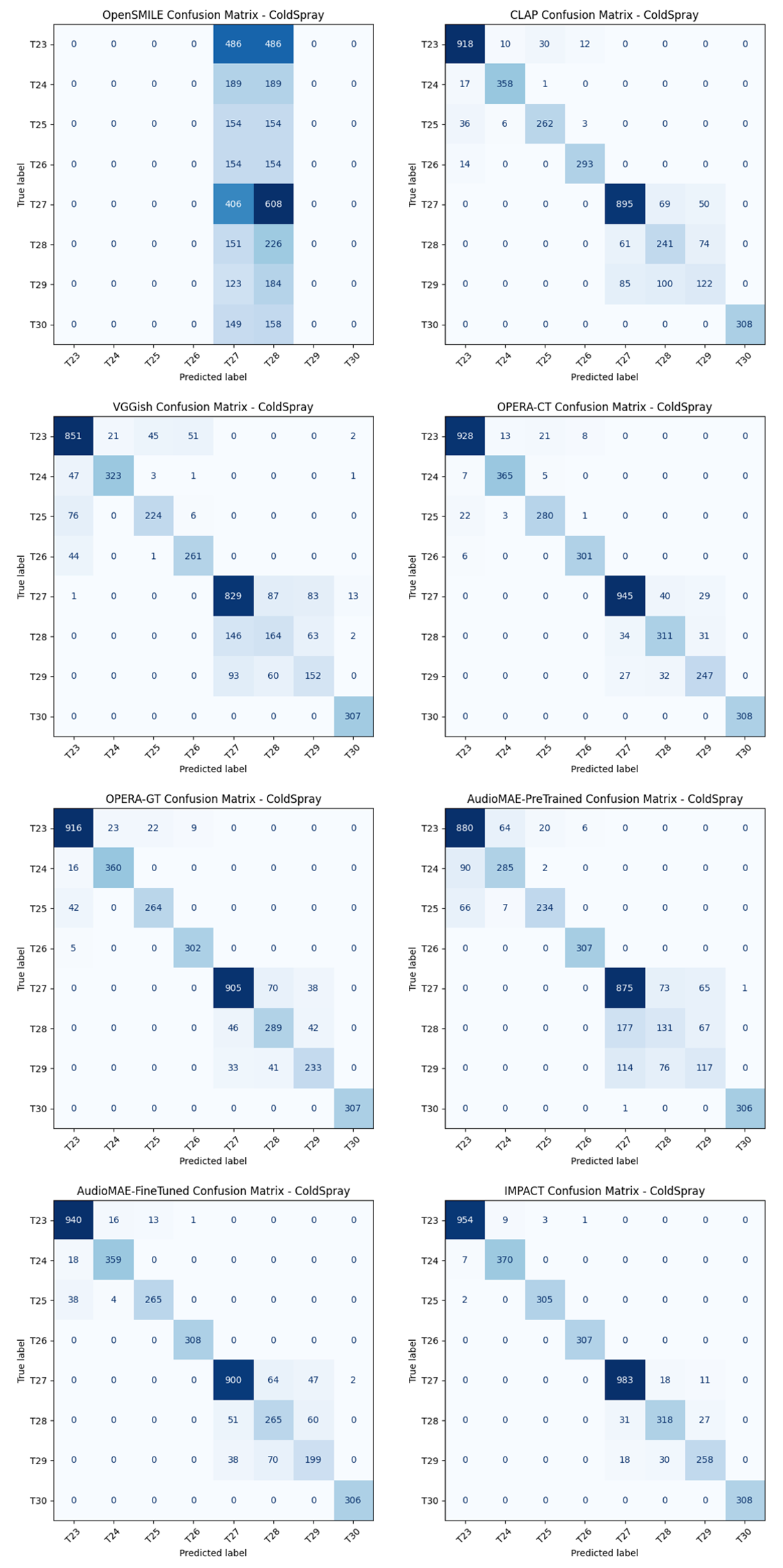}
            \caption{\textbf{Confusion Matrices for ColdSpray}. Each cell in the table shows the count of instances predicted by the model for the corresponding class.}
            \label{fig:cm_coldspray}
            \end{figure}

 Figure~\ref{fig:visualizedresults1} shows overall model performance per machine, while Figure~\ref{fig:visualizedresults2} present per-class performance across tasks T1–T30. The bars represent the average F1 scores, and the error bars indicate the standard deviations. IMPACT consistently achieves the highest performance to both intra-task complexity and inter-modality variation, demonstrating robust cross-machine generalization. Figure~\ref{fig:cm_renishawl},\ref{fig:cm_vf2},\ref{fig:cm_yornew},\ref{fig:cm_coldspray} represent the count of instances predicted by the model for the corresponding machine and class.
            
\newpage


\end{document}